\documentstyle[twocolumn,aps]{revtex}
\begin{document}
\input{epsf}
% \draft command makes pacs numbers print
\draft
%\preprint{Version 2.1}
%\vspace {5mm}
\title{Quantum spins mixing of spinor Bose-Einstein condensates}
\author{\\ {C.K. Law$^{1}$,  {H. Pu}$^{2}$ and N.P. Bigelow$^{2}$}}
\address{{$^{1}$Rochester Theory Center for Optical Science 
and Engineering} \\
{and Department of Physics and Astronomy,} 
{University of Rochester, Rochester, New York 14627}
\\
{$^2$Laboratory for Laser Energetics, and Department of Physics and 
Astronomy} \\
{University of Rochester, Rochester, New York 14627}}
%\date{\today}
\maketitle
%\vspace {28mm}
%\maketitle
%\vspace {28mm}
\begin{abstract}
We examine the internal structure of the ground states
of a trapped Bose-Einstein condensate in which atoms
have three internal hyperfine spins. We determine a 
set of collective spin states which minimize the 
interaction energy between condensate atoms. We also 
examine the internal dynamics of an initially spin 
polarized condensate. The time scale of spin-mixing 
is predicted.

\end{abstract}
\vskip 0.2in
\pacs{PACS numbers: 03.75.Fi }

\narrowtext

Bose-Einstein condensates (BEC) of atoms with internal 
degrees of freedom are new forms of macroscopically coherent 
matter which exhibit rich quantum structures. In the case of 
BEC with two internal spin states \cite{Myatt,Hall}, theoretical 
studies have predicted interesting phenomena such as quantum entanglement 
of spins \cite{Cirac}, suppression of quantum phase diffusion \cite{Law} 
\and interference effects \cite{Savage}. Recently, Stamper-Kurn {\it et al.}
\cite{Stamper} have realized an optically trapped BEC in which all three 
hyperfine states in the lowest energy manifold of sodium atoms are involved. 
Such a three-component condensate raises new questions regarding the more 
complex ground state structure \cite{Stenger,Ho} and internal spin dynamics. 
One of the key features here is that there are spin exchange interactions 
which constantly mix different condensate spin components while the system 
as a whole remains in the ground state. For example two atoms with respective 
hypefine spins $+1$ and $-1$ interact and become two atoms with hypefine spin 0. 
Therefore an important problem is to determine how atoms organize their spins 
in the ground state and how a spin polarized BEC loses its polarization 
because of spin exchange interactions. 
  
In this paper we approach the questions using an algebraic method 
found in quantum optics. We identify the fact that the interaction 
between spin components in a BEC is analogous to 4-wave mixing in nonlinear 
optics. However, since the trap is like a matter wave cavity, a more 
appropriate optical analogy is the 4-wave mixing in a high finesse 
cavity (i.e., a cavity QED system). With the help of the methods
developed in a related cavity QED problem \cite{Puri,Wu}, 
we are able to study the organization of spins in the condensate
ground state. We find that there exists a class of quantum superposition 
states which minimize the interaction energy. These quantum states are 
recognized as collective spin states which are characterized by strong 
correlations among different spin components, and in some cases we find 
that the number of atoms in individual spin component shows large fluctuations. 
In this paper we also examine the {\em internal dynamics} of the spin-mixing 
process arising from the nonlinear interactions between condensate atoms  
\cite{Goldstein}. For an initially spin polarized BEC, we predict the time 
scale at which spins become strongly mixed.

To begin we consider a dilute gas of trapped bosonic atoms with 
hyperfine spin $f=1$. The second quantized Hamiltonian of the system 
is given by $(\hbar =1)$,
\begin{eqnarray}
{\cal H}&=&\sum\limits_\alpha ^{} {\int_{}^{} {d^3x}\hat \Psi _\alpha ^{\dagger}
\left( {-{{\nabla ^2} \over {2M}}+V_T} \right)}\hat \Psi _\alpha  \nonumber \\
&+&\sum\limits_{\alpha ,\beta ,\mu ,\nu }^{} \Omega_{\alpha \beta \mu \nu }
{\int_{}^{} {\hat \Psi _\alpha ^{\dagger}
\hat \Psi _\beta ^{\dagger}\hat \Psi _\mu \hat \Psi _\nu d^3x}}
\end{eqnarray}
where $\hat \Psi_{\kappa}$ $(\kappa=-1,0,1)$ is the atomic field annihilation 
operator associated with atoms in the hyperfine spin state 
$\left| {f=1,m_f=\kappa } \right\rangle $. The summation indices
in (1) run through the values $-1,0,1$. The mass of the atom is given by
$M$ and the trapping potential $V_T$ is assumed to be the
same for all three components. The interactions between atoms are 
characterized by the coefficients $\Omega_{\alpha \beta \mu \nu }$ which are 
obtained from the two-body interaction model \cite{Ho,Goldstein,Ohmi,Walls},
\begin{equation}
U(\vec x_1,\vec x_2)=\delta (\vec x_1-\vec x_2)
\sum\limits_{F=0}^2 {g_F\sum\limits_{M_F=-F}^F {\left| {F,M_F} \right\rangle }}
\left\langle {F,M_F} \right|.
\end{equation}
Here ${\left| {F,M_F} \right\rangle }$ is the total hyperfine spin state
formed by two atoms each with spin $f=1$, and  
$g_F \equiv {{4\pi \hbar ^2a_F} / M}$ with $a_F$ is the $s-$wave scattering 
length in the $F$ channel. The interaction (2) is based on rather general 
symmetry assumptions of the system, because it preserves angular 
momentum and the rotation symmetry in hyperfine spin space \cite{Ho}. 
The model also makes use of the $\delta$ potential approach which has 
been widely used in one-component dilute BEC. 

By expanding the total spin state ${\left| {F,M_F} \right\rangle }$ 
in terms of basis vectors 
$\left| {f=1,m_f=\alpha } \right\rangle \otimes \left| {f=1,m_f=\beta } 
\right\rangle $, we obtain the Hamiltonian in the form   
${\cal H}={\cal H}_S+{\cal H}_A$, where 
\begin{eqnarray}
{\cal H}_S&=&\sum\limits_\alpha ^{} {\int_{}^{} {d^3x}
\hat \Psi _\alpha ^{\dagger}
\left( {-{{\nabla ^2} \over {2M}}+V_T} \right)}\hat \Psi _\alpha  \nonumber \\
&+&{\lambda _s \over 2}\sum\limits_{\alpha ,\beta }^{} {\int_{}^{} 
{\hat \Psi _\alpha ^{\dagger}\hat \Psi _\beta ^{\dagger}
\hat \Psi _\alpha \hat \Psi _\beta d^3x}}
\end{eqnarray}
\begin{eqnarray}
{\cal H}_A&=&{{\lambda _a} \over 2}\int_{}^{} 
{\left( {\hat \Psi _1^{\dagger}
\hat \Psi _1^{\dagger}
\hat \Psi _1\hat \Psi _1} \right.}
+\hat \Psi _{-1}^{\dagger}
\hat \Psi _{-1}^{\dagger}\hat \Psi _{-1}\hat \Psi _{-1} \nonumber \\
&+&2\hat \Psi _1^{\dagger}\hat \Psi _0^{\dagger}\hat \Psi _1
\hat \Psi _0+2\hat \Psi _{-1}^{\dagger}
\hat \Psi _0^{\dagger}\hat \Psi _{-1}\hat \Psi _0 \nonumber \\
&-&2\hat \Psi _1^{\dagger}\hat \Psi _{-1}^{\dagger}\hat \Psi _1\hat \Psi _{-1}
+2\hat \Psi _0^{\dagger}\hat \Psi _0^{\dagger}\hat \Psi _1\hat \Psi _{-1} \nonumber\\
&+& \left. {2\hat \Psi _1^{\dagger}
\hat \Psi _1^{\dagger}\hat \Psi _0\hat \Psi _0} \right)
\end{eqnarray}
Here $\lambda_s = (g_0 + 2g_2)/3$ and $\lambda_a = (g_2-g_0)/3$ are
defined. The Hamiltonian ${\cal H}$ is written as the sum of 
a symmetric part ${\cal H}_S$ and a non-symmetric part ${\cal H}_A$, where
${\cal H}_S$ remains unchanged for any interchange of the spin component 
indices.

In this paper we assume that the symmetric interaction ${\cal H}_S$ 
is strong compared with ${\cal H}_A$. This occurs for atoms whose 
scattering lengths in different $F$ channels have approximately same values 
such that $|\lambda_s| \gg  |\lambda_a|$. Recent estimations have 
indicated that sodium and rubidium atoms indeed have such a property.
With the symmetric ${\cal H}_S$ being the dominant Hamiltonian, 
the condensate wavefunctions $\phi_{\kappa}(\vec x)$ $(\kappa =0,\pm 1)$
for each spin component are approximately described by the same wavefunction
$\phi (\vec x)$, i.e., $\phi_{\kappa} (\vec x)=\phi (\vec x)$, which 
is defined by the Gross-Pitaevskii Equation through ${\cal H}_S$ ,
\begin{equation}
\left( {-{{\nabla ^2} \over {2M}}
+V_T+ \lambda_s N|\phi |^2} \right)\phi =\varepsilon \phi 
\end{equation}
where $\varepsilon$ is the mean field energy or the chemical potential.

Under the condition that atoms in different spin states are described 
by the same wavefunction, we can approximate field operators at the zero
temperature by,
\begin{equation}
\hat \Psi _\kappa 
\approx \hat a_\kappa \phi (\vec x)  \ \ \ \ \ \ \  \kappa=0,\pm 1
\end{equation}
Here $\hat a_{\kappa}$ is the annihilation operator 
associated with the condensate mode, and it satisfies the usual
commutation relation  $\left[ {\hat a_\kappa ,\hat a_\gamma } \right]=0$ and
$\left[ {\hat a_\kappa ,\hat a_\gamma ^{\dagger}} \right]
=\delta _{\kappa \gamma }$. Using (5) and (6), ${\cal H}_S$ and ${\cal H}_A$ 
have leading parts $H_s$ and $H_a$ respectively,
\begin{eqnarray}
{\cal H}_S & \approx & \varepsilon \hat N-\lambda _s' \hat N(\hat N-1)  
\equiv H_s\\
{\cal H}_A & \approx & \lambda _a'\left( {\hat a_1^{\dagger}\hat a_1^{\dagger}
\hat a_1\hat a_1+\hat a_{-1}^{\dagger}\hat a_{-1}^{\dagger}\hat a_{-1}\hat a_{-1}} \right. 
+2\hat a_1^{\dagger}\hat a_0^{\dagger}\hat a_1\hat a_0 \nonumber \\
&+&2\hat a_{-1}^{\dagger}\hat a_0^{\dagger}\hat a_{-1}\hat a_0
-2\hat a_1^{\dagger}\hat a_{-1}^{\dagger}\hat a_1\hat a_{-1}
+2\hat a_0^{\dagger}\hat a_0^{\dagger}\hat a_1\hat a_{-1} \nonumber \\
&+&\left. {2\hat a_1^{\dagger}\hat a_{-1}^{\dagger}\hat a_0\hat a_0} 
\right)\equiv H_A
\end{eqnarray}
Here $2\lambda _i' \equiv \lambda _i  \int_{}^{} {\left| {\phi (\vec x)} 
\right|^4d^3 x}$ $(i=s,a)$, and $\hat N \equiv \hat a_1^{\dagger}\hat a_1 
+\hat a_0^{\dagger}\hat a_0+\hat a_{-1}^{\dagger}\hat a_{-1}$ is the total
number of atoms in the condensate.

Our goal is to find the quantum states that minimize the 
energy $H_s+H_a$. Since $H_s$ is a function of $\hat N$ only, $H_s$ 
is a constant operator for a fixed number of atoms. Therefore it is 
sufficient to look for the ground state of $H_a$. It is quite remarkable 
that a similar structure of $H_a$ also appeared in nonlinear wave-mixing 
processes in cavity QED \cite{Puri}. We follow Refs.\cite{Puri,Wu} and 
identify the algebraic structure of the system. 
We notice that the
operators $\hat L_-\equiv \sqrt 2\left( {\hat a_1^{\dagger}\hat a_0
+\hat a_0^{\dagger}
\hat a_{-1}} \right)$,
$\hat L_+\equiv \sqrt 2\left( {\hat a_0^{\dagger}\hat a_1
+\hat a_{-1}^{\dagger}
\hat a_0} \right) $ and 
$\hat L_z\equiv \left( {\hat a_{-1}^{\dagger}\hat a_{-1}
-\hat a_1^{\dagger}
\hat a_1} \right)$
obey angular momentum commutation relations: 
$\left[ {\hat L_+,\hat L_-} \right]=2\hat L_z$ and
$\left[ {\hat L_z,\hat L_\pm } \right]=\pm \hat L_{\pm}$.
In other words, the operators $\hat L_{+}$, $\hat L_{-}$ 
can be interpreted as raising and lowering operators 
of a kind of `orbital angular momentum', and $\hat L_z$ 
is the `z-component' in the standard notations. From the
theory of angular momentum, $\hat L^2$ and $\hat L_z$ 
have a complete set of common eigenvectors 
$\left| {l,m_l} \right\rangle$ defined by
\begin{eqnarray}
&& \hat L^2\left| {l,m_l} \right\rangle=l(l+1)\left| {l,m_l} \right\rangle 
\\
&& \hat L_z\left| {l,m_l} \right\rangle=m_l\left| {l,m_l} \right\rangle
\end{eqnarray}
where $m_l=0,\pm 1, \pm 2,...,\pm l$. For a given total number of 
atoms $N$, the allowable values of $l$ are $l=0,2,4,...N$ if $N$ 
is even, and $l=1,3,5,...,N$ if $N$ is odd.
 
With the help of the angular momentum operators, $H_a$ takes
a very simple form,
\begin{equation}
H_a=\lambda _a'\left( {\hat L^2-2\hat N} \right).
\end{equation}
This is the main result of the paper because the
energy spectrum of $H_a$ is now solved. Eq. (11) 
indicates that $\left| {l,m_l} \right\rangle$ 
are eigenstates of $H_a$ with the energy $E_{l}^{a}$
\begin{equation}
E_l^{a} = \lambda _a'\left[ {l(l+1)-2N} \right].
\end{equation}
The lowest energy state of $H_a$ depends on the sign of $\lambda_a'$. 
In the following we discuss two cases: (I)  $\lambda_a' >0 $ and
(II) $\lambda_a' < 0 $.

\noindent (I) $\lambda_a' >0 $: In this case 
$\left| {l=0,m_l=0} \right\rangle $ is the ground state of $H_a$. 
Using the Fock states 
$\left| {n_1,n_0,n_{-1}} \right\rangle $ defined by the 
number operators $\hat n_j \equiv \hat a_{j}^{\dagger} \hat a_{j}$ 
for the three spin components (i.e., 
$ \hat n_j \left| {n_1,n_0,n_{-1}} \right\rangle 
=n_{j} \left| {n_1,n_0,n_{-1}} \right\rangle$),
$\left| {l=0,m_l=0} \right\rangle $ has the form
\begin{equation}
\left| {l=0,m_l=0} \right\rangle 
=\sum\limits_{k=0}^{[N/ 2]} {A_k\left| {k,N-2k,k} \right\rangle }
\end{equation}
where the amplitudes $A_k$ obey the recursion relation
\begin{equation}
A_k=-\sqrt {{{N-2k+2} \over {N-2k+1}}}A_{k-1}.
\end{equation}
We see that the state $\left| {l=0,m_l=0} \right\rangle$ is 
a quantum superposition of a chain of Fock states 
$\left| {k,N-2k,k} \right\rangle$ in which the numbers of atoms 
in the spins $1$ and $-1$ are equal. We stress that such a quantum 
state is a {\em collective} spin state which cannot be expressed as 
product states of individual atoms. The amplitudes $A_k$ are
arranged in such a way that the interaction energy
$H_a$ is almost completely cancelled. This can be seen
from the disappearence of $N^2$ dependence in the
energy of $H_a$.  It is not difficult to show that for
the state (13), the average numbers of atoms in each component 
are all equal, i.e., $\left\langle {\hat n_0} \right \rangle 
= \left\langle {\hat n_1} \right \rangle 
=\left\langle {\hat n_{-1}} \right \rangle =N/3$.
Since $A_k$ are almost uniformly distributed (see Fig. 1a), there are 
large fluctuations of particle numbers in individual components although 
the total particle number $N$ is fixed. More precisely, we find that
$\left\langle {\Delta \hat n_0} \right \rangle \approx 2N/ \sqrt{5}$
for $N \gg 1$, i.e., a super-Poisson distribution. Our further calculations 
indicate that super-Poisson distribution of particle numbers are a common 
feature for low energy eigenstates of $H_a$ when $\lambda_a' >0 $.

\noindent (II) $\lambda_a' <0 $: In this case $H_a$ has 
$2N+1$ degenerate ground states given by $\left| {l=N,m_l} \right\rangle $ 
where
$m_l=0,\pm 1, \pm 2,...,\pm N$. The energy (12) of these states 
is $\lambda _a'N(N-1)$, and the general form of 
$\left| {l=N,m_l} \right\rangle $ is given by
\begin{equation}
\left| {l=N,m_l} \right\rangle 
=\sum\limits_k^{} {B_k^{(m_l)}\left| {k,N-2k-m_l,k+m_l} 
\right\rangle }.
\end{equation}
Here the summation index $k$ runs over all physical Fock
states $\left| {k,N-2k-m_l,k+m_l} \right\rangle$ (i.e., those with 
non-negative numbers in each component). The simplest case of (15) is 
$\left| {l=N,m_l=-N} \right\rangle = \left| {N,0,0} \right\rangle$,
and with this we can construct the amplitudes $B_k^{(m_l)}$ by 
repeatedly applying the raising operator $\hat L_+$.
To give an illustration, we plot in Fig. 1b the Fock state 
amplitudes $B_k^{(m_l)}$ for several $m_l$'s.
We see that $B_k^{(m_l)}$ has a narrow distribution which 
indicates well defined particle numbers in each spin components. 
It is interesting that all the degenerate states (15) have sub-Poisson 
number fluctuations in each spin component (see the inset of Fig. 1b). 
This feature is just the opposite to the previous case $\lambda_a' >0 $.

Finally, let us look at the spin-mixing dynamics
of an initially spin-polarized condensate in which
all atoms in the condensate are prepared in the 
spin 0 state at $t=0$, i.e., $
\left| {\psi (0)} \right\rangle =\left| {0,N,0} \right\rangle $.
In this case two atoms in the spin 0 state can be converted
into one atom in the spin $1$ state and the other in 
the spin $-1$ state. Using the $H_s+H_a$ as our 
approximate Hamiltonian, the system at time $t$ is given by,
\begin{equation}
\left| {\psi (t)} \right\rangle 
=e^{-i\theta_N (t) }\sum\limits_{l=0}^{N} 
{C_l}e^{-i\lambda _a'l(l+1)t}\left| {l,m_l=0} \right\rangle 
\end{equation}
where $C_l=\left\langle {{l,m_l=0}} 
\mathrel{\left | 
{\vphantom {{l,m_l=0} {0,N,0}}} \right. 
\kern-\nulldelimiterspace} {{0,N,0}} \right\rangle $
and $\theta_N (t)  = (\varepsilon N+\lambda _s'N(N-1))t$.
In Fig. 2 we present the time dependence of the particle number 
in the spin-0 component for $N=10^2,10^3,10^4$ cases. We see that 
the number of atoms in the spin-0 component becomes steady 
at $\left\langle {\hat n_0} \right\rangle=N/2$ 
after a time $t_c$,
\begin{equation}
t_c \approx {1 \over {2\lambda_a' \sqrt{N}}}.
\end{equation}
This is the time scale for the spin-mixing process purely
due to the nonlinear interaction between condensate atoms \cite{remark}.
In the Thomas-Fermi (large $N$) limit, we find that for a spherial 
harmonic trap, $(g_2-g_0)t_c\approx 5.1 N^{1/ 10}
\left( {(g_0+2g_2) / M\omega ^2} \right)^{3/ 5}$ where $\omega$
is trap frequency. Therefore $t_c$ becomes quite insensitive to $N$ 
in the Thomas-Fermi limit. To give a realistic example, for a 
sodium condensate with $N= 10^4$ and $\omega =2\pi \times$370Hz, 
we find that $t_c$ is about $0.4$ seconds, assuming $a_2 \approx 2.6$nm
and $a_0 \approx 2.3$nm. 

We remark that the spin mixing dynamics can be quite different for 
different initial conditions. In Fig. 3, we give an example for the case 
when all three components initially have the same atom numbers, i.e.,
$\left| {\psi (0)} \right\rangle =\left| {N/3,N/3,N/3} \right\rangle$. 
It is quite surprising that the particle number 
executes fast oscillations with a frequency of the order of 
$\lambda_a' N$, and then the system suddenly becomes steady. 
This interesting behavior indicates that there are complex
quantum dynamics governed by the nonlinear interaction $H_a$.
In fact since $H_a$ has a discrete spectrum, quantum recurrence or 
revival is expected in a much longer time scale
(which is typically of the order of $\pi / \lambda_a'$). It is worth 
further exploring the quantum dynamics in the context of either 
BEC or cavity QED.

To conclude, we have examined the spin-mixing interaction
of a Bose-Einstein condensate with three internal spin
components. It is quite remarkable that the model
interaction (2) (which is based on general symmetry assumptions)
can lead to a simple algebric representation, and from which 
we can construct the collective spin states which minimize 
the interaction energy among condensate atoms. These collective 
states exhibit spin correlations and characteristic
particle number fluctuations which depend crucially on the 
sign of $\lambda_a'$. We have also investigated the spin mixing dynamics 
due to the nonlinear interaction between condensate atoms. 
The time scale of mixing for an initially spin polarized system 
is identified. This study provides a theoretical treatment 
of the structure and dynamics of spinor BEC. However, our analysis 
are limited to interaction between condensate atoms, it remains
to be answered that how non-condensate atoms will decohere the 
condensate structure. Further work along this direction would be
necessary. 

\acknowledgments
CKL would like to thank Prof. J.H. Eberly for discussions. 
This research was supported by NSF grants PHY-9415583 and PHY-9457897, 
and the David and Lucile Packard Foundation.

\vspace{-20mm}
\begin{figure}
\centerline{
\epsfxsize=3.5in
\epsfbox{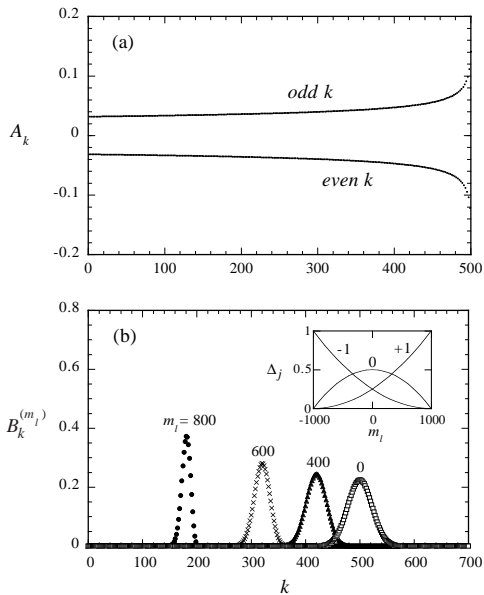}}
\vspace{-10mm}
\caption{Amplitudes of Fock states associated with the ground
states of $H_a$ for $N=10^3$ atoms: (a) $\lambda_a' >0$, (b) 
$\lambda_a' <0$. The inset in (b) shows the normalized number 
fluctuations 
$\Delta _j\equiv \left\langle {\Delta n_j} 
\right\rangle ^2 / \left\langle {\hat n_j}  
\right\rangle $  in the three spin components $(j=0,\pm 1)$ 
as a function of $m_l$. Sub-Poisson distributions are defined 
by $\Delta_j < 1$.}
\end{figure}

\vspace{-20mm}
\begin{figure}
\centerline{
\epsfxsize=3.5in
\epsfbox{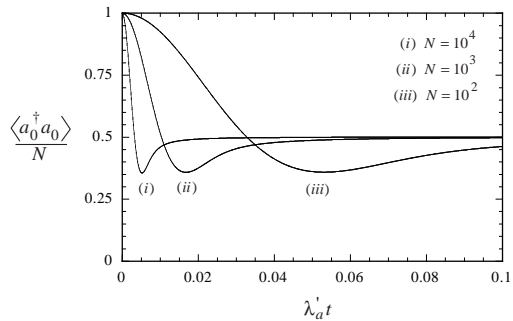}}
\vspace{-20mm}
\caption{Time dependence of average number of atoms in the spin 0 state
normalized by the total number of atoms $N$. The initial state of the 
system is $\left| {\psi (0)} \right\rangle =\left| {0,N,0} \right\rangle$. 
We show three cases with $N=10^2,10^3, 10^4$.}
\end{figure}

\vspace{-20mm}
\begin{figure}
\centerline{
\epsfxsize=3.5in
\epsfbox{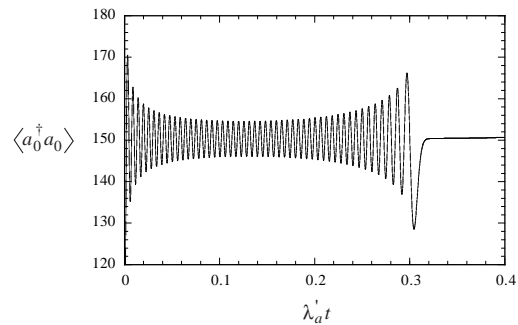}}
\vspace{-20mm}
\caption{Time dependence of average number of atoms in the spin 0 state. 
The initial state of the system is 
$\left| {\psi (0)} \right\rangle =\left| {N/3,N/3,N/3} \right\rangle$,
where $N=300$.}
\end{figure}

\narrowtext


\begin{references}
\bibitem{remark} Our further calculations indicate that 
$\left\langle {\Delta \hat n_j } \right\rangle$
also become steady for $t >t_c$. These particle number fluctuations 
(for $t>t_c$) are found to be quite significant:
$\left\langle {\Delta \hat n_1} \right\rangle 
= \left\langle {\Delta \hat n_1} \right\rangle
= \left\langle {\Delta \hat n_0} \right\rangle /2 
\approx 0.18 N$.
\end{references}
\end{document}